# Microscopic Spectral Model of High Temperature Superconductors

J. C. Phillips

Dept. of Physics and Astronomy, Rutgers University, Piscataway, N. J., 08854-8019

**Abstract**

The self-organized dopant percolative filamentary model, entirely orbital in character (no spins), explains the evolution with doping of Fermi arcs observed by ARPES, including the abrupt transitions in quasiparticle strength observed near optimal doping in cuprate high temperature superconductors. Similarly abrupt transitions are also observed in time-resolved picosecond relaxation spectroscopy at 1.5 eV, and these are explained as well, using no new assumptions and no adjustable parameters. The anomalous "precursive" temperature-dependent strains observed by EXAFS are associated with relaxation of filamentary ends.

## 1. Introduction

High temperature superconductors exhibit many anomalous properties, as revealed in a wide variety of experiments. Spectroscopic studies, particularly by angle-resolved photoemission (ARPES), have shown many surprising features of states near the Fermi energy, that correlate in entirely unexpected ways with the superconductive phase diagram. The latter characteristically exhibits three phases, an insulating phase (dopant concentration x in the range $0 < x < x_1$), an intermediate superconductive phase with anomalous normal-state transport ($x_1 < x < x_2$), and a non-superconductive metallic or Fermi-liquid phase ($x > x_2$) [1]. The superconductive transition temperature $T_c(x)$ reaches a maximum near the center of the intermediate phase at the optimal doping $x = x_0$. Yet ARPES experiments [2] on $La_{2-x}Sr_xCuO_4$ (LSCO, $x_1 = 0.06$, $x_2 = 0.22$, $x_0 = 0.16$) showed metallic behavior in the Fermi arc even at $x = 0.03$ (insulating phase) beginning in the nodal $(\pi,\pi)$ gap direction, and spreading smoothly over a wider angle as x increased. Moreover, the intensity of metallic signal exhibited an abrupt qualitative change at $x = x_0$, with the largest value shifting from the nodal $(\pi,\pi)$ gap direction to the antinodal $(\pi,0)$ gap direction for $x > x_0$.

Spectroscopic experiments involve averages over lengths large compared to atomic dimensions, and so apparently contain information describable entirely in terms of order parameters based on mean field theory. Order parameter descriptions also have the advantage of being non-committal at the microscopic level, but at the same time are uninformative as to the microscopic mechanisms giving rise to the observed anomalies. By contrast, large area, high-resolution scanning tunnel microscopy (STM) studies [3] have exhibited microscopically inhomogeneous nanodomain gap structure that also varies smoothly with x. Moreover, these studies showed that Fourier transforms of the gap structure reproduce the d-wave gap order parameter observed by ARPES; at the same time, they also showed that the d-wave part of the gap distribution was anomalously small (too small to be metallic in a randomly disordered medium). These paradoxical features of the phase diagram and both macroscopic and microscopic electronic features lead to a glassy, self-organized (fully non-random, stronger than directed, fully adaptive) percolative model based on self-organized *dopant* filaments [4]. The model explains the strongly counter-intuitive x dependence of the isotope shifts observed [5] both above and below the ARPES phonon kinks [6].

The isotope shifts unambiguously identify electron-phonon interactions as the microscopic origin of high temperature superconductivity, with the counter-intuitive behavior explained by the difference between the free (intuitive) quasiparticle behavior of a gas and the nearly frozen (counter-intuitive) behavior of a dopant glass. Recent improvements in ARPES resolution have even led to identification of phonon fine structure in the nodal directions in LSCO [7] similar to the Eliashberg fine structure occasionally seen much earlier in tunneling experiments [8].

Phonon interactions occur on a picosecond time scale, and recent time-resolved studies of reflectance R relaxation in $Bi_2Sr_2CaCu_2O_{8+x}$ (BSCCO) showed dramatic x dependencies of $\Delta R/R$ and relaxation time $\tau$ on that time scale with a pump-probe energy of 1.5 eV [9], that closely parallel the ARPES x dependencies, with an abrupt change in $\tau(x)$ at $x_0$, and a more gradual sign reversal of $\Delta R(x)/R$ between $x_1$ and $x_2$, passing through 0 at $x = x_0$. [9] discuss their data in terms of continuum concepts (such as weak localization) whose validity has been established mainly in pure, undoped Fermi liquid metals (such as Ag) that are not superconductive at high temperatures and do not exhibit intermediate or insulating phases. Here we show that the glassy, self-organized percolative model [4] based on self-organized *dopant* filaments describes both the ARPES and time-resolved reflectance data without using new or additional assumptions.

2. **Elements of the topological model**

The modern theory of glasses is called constraint theory [1,6,10]; it describes accurately the phase diagrams of both molecular and electronic glasses. In the molecular case the constraints are counted by analyzing interatomic bonding forces (stretching, bending…). The mean-field condition for forming an ideal glass is $N_c = N_d$, where $N_c$ is the number of constraints per atom, and $N_d$ is the number of degrees of freedom per atom. In a gas of N atoms, $N_d = 2(N - d)/N$, but in a glass $N_d = d$, as the glass is frozen into a configuration of nearly maximal density, with only spatial degrees of freedom (periodic crystalline configurations are, of course, not glassy). Cuprates are electronic glasses, with the dopants frozen into configurations that nearly optimally screen internal electric fields. Constraint theory is topological, not analytical. Analytical models of continuum systems have been able to derive results not only for simple nearly free electron metals, but even for weakly disordered metals (dirty transition metals, …). They are, however, completely unsuited to strongly disordered, nanoscopically inhomogeneous metals with anomalous properties obtained by adaptively doping an insulator, such as the cuprate high temperature superconductors [3].

While there are many crystalline materials, there are only a few ideal glass formers. These in turn have properties that are very, very different from those of normal materials, and these properties are often of great value, which has focused attention on them. Constraint theory correspondingly is of little value in discussing most materials, because the ideal glass-forming condition $N_c = N_d$ is seldom satisfied. However, when circumstances make it possible to form ideal glasses, constraint theory is very powerful [1]. It is able to explain, without using adjustable parameters, why cuprates (and to a lesser extent the $(Ba,K)(Pb,Bi)O_3$ family) are the only materials that form high temperature superconductors [11], a highly material-specfic result inaccessible to analytic theories, even with adjustable parameters. Similarly it explains [6] the counter-intuitive isotope effects, again without using adjustable parameters, and again a highly material-specific result inaccessible to analytic theories, even with adjustable parameters.

Fig. 1 illustrates the basic topological model for layered cuprate high temperature superconductors. Semiconductive layers (such as SrO) alternate with metallic layers ($CuO_2$, BiO). The planar lattice constant is fixed by the $CuO_2$ layers, which stabilize the structure mechanically and satisfy the ideal glass-forming condition $N_c = N_d$ with regard to interatomic forces [11]. The remaining layers are softer and are easily distorted: this explains why the only electronic states near $E_F$ that are observable by ARPES are those associated with $CuO_2$ layers. It is customary to assume that the $CuO_2$ layers are perfectly crystalline, but because of the interlayer misfit to semiconductive layers of effectively different relaxed lattice constants, this is almost certainly not the case [10,11,12]. The misfit can be relieved, and the total energy reduced, by inserting thin insulating domain walls in the $CuO_2$ layers; the natural choice for such walls would be doubled unit cells, distorted in such a way as to introduce a small energy gap, analogous to that formed by a charge density wave. Because of the overall softness of the lattice, the gain in electronic energy from forming such a pseudogap is reduced only partially by the increased elastic energy.

The splitting of the metallic planes into nanodomains separated by thin insulating domain walls results in a narrow gap semiconductor. Now as the x dopants are added, usually to the alternating semiconductive layers (such as SrO), they provide bridges that can carry current in zigzag paths around the domain walls [4]. These paths form dopant-centered conducting wires, and there is an insulator-metal transition when there are enough dopants that the wires begin to percolate at $x = x_1$. This percolative metal phase is qualitatively different from a quasi-one-dimensional normal metal with weak localization, because the filamentary density increases with x. At $x = x_0$ it reaches a maximum, and at larger values of x the filaments bunch together so closely that interfilament scattering converts the bunched regions into regions that are locally normal metals [6]. In these normally metallic regions the local density of states near $E_F$ is much larger than in the filaments. These zigzag paths are a useful tool for analyzing the relation between anomalies in LO phonon spectra measured by neutron scattering [13]. They also provide a derivation [14] of the d-wave gap by projection of the paths on preferred antinodal directions; in this microscopic model nodal gap quasiparticles are replaced by *projections* of Cooper pairs bound to independent filaments onto nodal $(\pi,\pi)$ gap directions; these are the directions along which percolation between nanodomains with edges parallel to Cartesian directions are most effective.

A more formal description of the origin of the zigzag paths proceeds as follows. For simplicity we assume that all the dopant bridges are locally equivalent. (This is a good approximation because the domain walls are similar; they relieve the global interlayer misfit. Similarly, the dopants are similar, because they adopt the optimal bridging position that maximizes the dopant-assisted tunneling through (or around) the dopant walls. In a mean field approximation the coherence of the dopant impurity band network is now measured by the structure factor $S(\mathbf{k}) = \Sigma_i \exp(i\mathbf{k}\cdot\mathbf{r}_i)$, where the sum extends over all dopant sites $\mathbf{r}_i$. Such a sum, in the presence of many-electron thermal fluctuations, rapidly becomes incoherent. However, if we separate the sum into its filamenatary components as a double sum $S(\mathbf{k}) = \Sigma_f \Sigma_i \exp(i\mathbf{k}\cdot\mathbf{r}_{if})$, then each filament contain only a few electrons, so that the average interlevel energy spacing on a given filament (now of order $N^{-1/3}$ instead of $N^{-1}$) becomes large compared to kT. Thus the filaments are separately coherent or incoherent (each filament has its own phase or order parameter), and the coherence of the "intact" filaments is affected only marginally by interfilamentary interactions with the incoherent filaments.

## 3. ARPES Arcs

By now it is well known that Fermi liquid (mean field) theory does not provide an adequate description of the electronic spectra of cuprate high temperature superconductors measured by ARPES. There are many difficulties: the Fermi arc does not shift with doping, but instead even in the insulating phase with $x = 0.03 < x_1 = 0.06$, states begin to appear at $E_F$ in the nodal directions [2], and with increasing x an occupied arc $\mathbf{k}(E_F)$ centered on this direction grows in width and intensity $Z(x,\alpha,E_F)$, where the angle $\alpha = \pi/4$ (0) for the gap nodal (antinodal) directions. The growth is not completely smooth (there are jumps in $Z(x,\pi/4,E_F)$ when x crosses $x_1$ and between $x = 0.10$ and $0.15$, Fig. 2(a)), but the most striking feature of this Fermi arc is a very large jump in $Z(x,0,E_F)$ when x crosses $x_0$ (goes from underdoped to overdoped, and $T_c$ goes through its maximum; see Fig. 2(b)). Because it is able to explain both the origin of the intermediate phase and the origin of d-wave gap anisotropy, topological constraint theory is ideally suited to discussing the abrupt transition in $Z(x,0,E_F)$ for $x = x_0$ observed in ARPES data [2]. For the reader's convenience the main features of these data are sketched in Fig. 2. (The ARPES data are quite complex and should be viewed in full color [2].)

The topological explanation for these unprecedented phenomena is that the highly mobile dopants self-organize to form conductive (electronically coherent) segments even in the insulating phase when annealed at high temperatures. By forming such conductive segments the dopants lower the sample free energy by better screening internal electric fields in this strongly ionic material. The anisotropy of the energy gap tells us that the antinodal (nodal) directions $\alpha = 0$ ($\pi/4$) are the directions of strongest (weakest) electron-phonon interactions. Thus the nodally oriented segments have the lowest scattering and the highest conductivity, and the segments that are formed at small $x = 0.03$ are therefore oriented predominantly in the $\alpha = \pi/4$ direction. With increasing x the segments begin to percolate at $x = x_1$ to form a metallic network. Because of the presence of a high density of insulating islands (large pseudogaps [3]) the percolative paths become sinuous, and the occupied Fermi surface arc spreads to larger values of $|\alpha - \pi/4|$ with increasing x. Note that these filamentary states are states bound or pinned to the dopants. Their momentum spectrum remains that of the metallic patches in the metallic planes, which is why the Fermi arcs do not move in **k** space as x increases and dopants are added to insulating layers. Note that the filamentary paths are three-dimensional, and are not confined to the metallic planes; when projected onto these planes, the filamentary states lie in the energy gaps of the nanodomain walls. Without <u>these</u> walls the filamentary bound states could not be formed.

Turning now to the most striking feature of the ARPES data, the very large jump in $Z(x,0,E_F)$ when x crosses $x_0$ (see Fig. 2(b)), the authors [2] explain this anomaly as arising from a flat energy band arising from states near $(\pi,0)$. However, it is not easy to see how such an energy band could be derived from a periodic potential. It will generate a very large peak in $N(E)$, the density of quasi-particle states, at $E = E_F$. Normally such a peak will split into two peaks, above and below $E_F$ (Jahn-Teller effect). Moreover, it would have to be narrow on an energy scale of order 0.01 eV. Within our model such states are easily pinned in the nanodomain wall energy gap, but without such a gap it seems to be impossible to generate such a flat band.

The flat band of states pinned to $E_F$ at $x = x_0$ must also consist of states localized in **k** space near $(\pi,0)$. In a band model these extra states would presumably be associated with a saddle point in k space located at $(\pi,0)$, but such a model cannot account for the abruptness of its appearance. In our model it is easy to see how this happens. As x increases towards optimal doping $x = x_0$, the weak scattering directions near $\alpha = \pi/4$ on the Fermi arc are filled by dopants with local filamentary tangents with α oreintations, and finally only the strong scattering directions near $\alpha = 0$ are left; these are filled last. Beyond $x = x_0$, additional dopants must occupy sites close enough to other dopants that strong scattering occurs at these dopant pairs. This strong scattering locally destroys filamentary coherence, giving rise to Fermi liquid (incoherent) patches. Because the most highly conductive directions are the $\alpha = \pi/4$ directions, the internal fields will be best screened by orienting the overdoped doping pairs along $\alpha = 0$ directions. The structure factor $S_2(\mathbf{k})$ of these closely spaced pairs then reflects the "flatness" (in **k** space) of the states responsible for the abrupt jump in $Z(x,0,E_F)$ when x crosses $x_0$ (see Fig. 2(b)). The dopant pairs are larger than the individual (atomic scale) dopants that form the underdoped filaments responsible for the nodal arc; the natural length scale for these pairs is the nanodomain diameter, of order 6 unit cells [3]. The structure factor of the overdoped pairs is thus ~ 6 times more localized in **k** space than the underdoped nodal structure. Also one should note that in the present model there is a sum rule: $\int d\alpha\, Z(x,\alpha,E_F) \sim x$. (To implement this rule one should include the dependence of state broadening $\Gamma(x, \alpha)$ in estimating Z from the data.) Thus the abrupt increase of $Z(x,0,E_F)$ when x crosses $x_0$ takes place at the expense of $Z(x,\alpha,E_F)$ for values of $\alpha$ not close to 0.

While the foregoing discussion justifies the usage of the term "flat bands" to explain the abruptness of the appearance of the strong intensity of incoherent antinodal states beyond $x = x_0$, it is important to realize that these states are not one-electron band states in the usual sense of rigid or fixed energy bands. Dopants close enough together to form Fermi liquid droplets develop quadrupole moments with the long axis oriented along the Cartesian axes; these clusters are many-electron systems, characterized by strongly anisotropic one-electron scattering. As soon as such droplets begin to form, $T_c$ begins to decrease, as filamentary coherence is disrupted by Fermi liquid incoherence. The (in)coherence arises from the local topology of the dopant clusters, and it is not a continuum property, even though it may be probed by photons with wave lengths large compared to the dopant spacing.

**4. Comparison with other theories**

Unlike analytical models, topological models are not adorned by elaborate algebra, but they also contain no adjustable parameters. Their great strength lies in their ability to identify and even predict the essential qualitative features of complex problems; here these problems arise from the need to identify the essential features of many-body dopant configurations relelvant toa given experiment. The Fermi arcs identified by ARPES have attracted great theoretical interest, previously discussed using analytical models; it is therefore useful to compare these analytical models with the present topological model, particularly with respect to qualitative features. Although by now it is abundantly obvious that electron-phonon interactions cause high temperature superconductivity, and that gas- or liquid-like quasiparticle models cannot account even qualitatively for the dependence of the phonon kink on x and isotopic mass [5,6], almost all the mean field models that claim to explain the Fermi arc in underdoped cuprates invoke spin (not phonon) degrees of freedom, usually in the context of multiply parameterized t-J Hubbard models with generous helpings of uncontrolled renormalization of two-dimensional energy gaps by three-dimensional interactions (another adjustable parameter) [15]. Strong claims are made for these models [16,17]: they contain the "essential" physics of the cuprates, the momemtum dependence of their quasiparticle states is "essentially" that observed by ARPES, etc. Similarly strong claims have been made for Fermi liquid models with strong spin relaxation channels [18].

A striking feature of all the spin-dependent mean field models is that they purport to explain optical data (which reflect almost entirely electric dipole transitions, with oscillator strengths determined by orbital coherence, in terms of both spin and orbital degrees of freedom, yet they never mention the awkward fact that the atomic spin-orbit coupling parameters $\lambda$ for O 2p and Cu 3d states are ~ 0.01 eV, which is negligibly small compared to the orbital band widths W ~ 1 eV. In these "strong coupling" Hubbard mean field models the spin and orbital degrees of freedom are coupled as artifacts of projection procedures driven by the Coulomb repulsion parameter U, assumed to be ~ 10W. These "central cell" (short range) projection procedures involve many uncontrolled approximations that are probably incompatible with coherent intercellular (long range) orbital motion, and the gradual appearance of the Fermi arc may be merely a result of the projection procedures used. (Note that in the parallel classical problem of mechanical elasticity phase changes from floppy to rigid, in the mean field approximation *there is no intermediate phase* [19]. It seems likely that were the quantum models to be treated exactly in mean field, the intermediate phase would disappear there as well; it appears as an artifact derived from smoothing out a first-order mean field phase transition.) In particular, there appears to be no way that these *smooth* projection procedures can generate the *abrupt* transition in $Z(x,0,E_F)$ when x crosses $x_0$ at optimal doping observed in ARPES (Fig. 2(b)).

## 5. Picosecond reflectance relaxation

The emphasis in the present self-organized (adaptive, not random) percolation model on formation of dopant filaments apparently requires a "Maxwell demon" to identify such dopants, especially as optical experiments involve wave lengths much longer than the nanodomain dimensions identifiable only by large area, high-resolution scanning tunnel microscopy (STM) studies [3]. However, time-resolved relaxation studies have proved to be a powerful tool in studying network self-organization in both molecular and electronic glasses [20,21], and elegant pump-probe reflectance relaxation studies [9] at 1.5 eV have revealed picosecond anomalies in $La_{2-x}Sr_xCuO_4$ (LSCO, $x_1 = 0.06$, $x_2 = 0.22$, $x_0 = 0.16$) that closely parallel the ARPES anomalies, with an abrupt change in relaxation time at optimal doping.

It seems surprising that a simple reflectance measurement, which averages over all **k** values involved in interband optical transitions, can yield an anomaly very similar to that obtained by the much more refined and precise ARPES **k**-resolved technique; this is possible because there are some very subtle aspects to pump-probe time-resolved relaxation studies in self-organized glassy systems [20,21]. These are illustrated in Fig. 1(c) and (d). The electron-hole pair created by photon absorption relaxes rapidly (femtosecond time scale) to a metastable excitonic state which then relaxes much more slowly, probably non-radiatively (phonon emission), on a picosecond time scale. The experimental relaxation data [9], reproduced for the reader's convenience in Fig. 3(a),(b), refer to this second regime. There are not one, but two surprises in these data. First, $\Delta R(x)/R$ smoothly reverses sign at $x = 0.16 = x_0$, and second, the relaxation time $\tau$ abruptly drops from 20 ps in the underdoped regime $x < 0.16 = x_0$ to $\sim 2$ ps for $x \geq 0.16$. As emphasized in [9], this is strongly reminiscent of the abrupt crossover in $Z(x,\pi/4,E_F)$ - $Z(x,0,E_F)$ in Fig. 2(b). At the same time, it is hard to understand how this abrupt drop can be associated with a peak in a quasiparticle density of states $N(E)$ pinned to $E_F$ at $x = x_0$.

(In a rigid band quasiparticle model $\tau$ would reach a minimum at $x = x_0$ and then recover for both $x < x_0$ and $x > x_0$.) The self-organized (not random) filamentary percolation model, with its strong local field corrections, readily explains both of these results (surprising and *basically inexplicable in a mean field quasiparticle context*).

Because the pump and probe both have energy 1.5 eV, this experiment detects changes that occur because of addition of a metastable exciton.to the filamentary network. The exciton is metastable because it has been added to a semiconductive nanodomain wall (excitons added to the metallic nanodomains decay on the femtosecond time scale), and it decays by "leaking out" (tunneling) to the metallic regions. The positions occupied by the metastable excition are not all equivalent, even though the probed single-particle exciton energy remains fixed at the pump energy 1.5 eV. The exciton is highly polarizable, and the free energy of the filamentary network depends on the position of the exciton. For $x < x_0$, the exciton diffuses to positions that minimize the system's free energy by increasing the conductivity of the filamentary network. Thus the exciton functions as a surrogate dopant, increasing the length and connectivity of coherent filaments, and hence increasing the reflectivity, $\Delta R/R > 0$. However, for $x > x_0$, the filaments have already filled the available phase space (optimized glass networks are "incompressible"), and when excitons are added, they function again as dopants, but this time as excess dopants, disrupting network segments and producing incoherent Fermi-liquid like regions.

There are two reasons why $\Delta R(x)/R$ changes smoothly with x. First, the high conductivity sites, with $\alpha = \pi/4$ filamentary tangents, are occupied first, and the lower conductivity sites (increasing $|\alpha - \pi/4|$) are then filled with increasing x. Second, note that for $x < x_0$, the exciton must diffuse from its initial region to find the best filamentary site, and that this is an easier task for strong underdoping, when the filaments are nearly oriented in $\alpha = \pi/4$ channels, than near optimal doping, when the filaments have become sinuous, and the entire network must reconfigure itself to improve its coherent conductivity. For $x > x_0$, the disruption is minimized for surrogate insertion near $\alpha = 0$ oriented filamentary tangents, where the local conductivity is less than in $\alpha = \pi/4$ channels. That is why the magnitude of $\Delta R/R$ is smaller for strong overdoping than for strong underdoping. One can add that a semiclassical simulation of coherent filamentary percolation could be carried out in the context of directed percolation, with angular weighting factors. Such a simulation would be considerably more complex than the mechanical simulations of [19], and it lies outside the framework of this paper.

The situation for $\tau(x)$ is different. For $x < x_0$, decay of the exciton must occur through a Franck-Condon configurational barrier (the exciton abandons its surrogate coherent filamentary function, and decays into an incoherent state), which leads to a large $\tau$. For $x > x_0$, the exciton decays from an incoherent state in the semiconductive wall to another incoherent state in the adjacent metallic nanodomain. Because both states are incoherent, the Franck-Condon configurational barrier almost disappears, and the decay takes place on a ps (one phonon) time scale (ten times faster than for $x < x_0$).

The measured abrupt drop in the average value of $\tau$ by a factor of 10 at $x = x_0$ can be estimated as follows. According to the discussion at the end of Section 2, in the metallic region $x > x_0$, the energy level spacing in a nanodomain is of order $W/N$, where $W \sim 1$ eV is the valence band width, and $N$ is the number of unit cells in the nanodomain. In the filamentary region $x < x_0$, the intrafilamentary energy level spacing in a nanodomain is of order $W/N^{1/3}$, so it is much smaller. All other things being equal, Fermi's golden rule for transition rates says that the ratio of the relaxation times will scale as $N^{2/3}$. This gives $N \sim 30$, which is consistent with the observed nanodomain dimensions of 3 nm [3].

## 6. Strain

Many other data indirectly show evidence with increasing x for the crossover from filamentary formation ($x < x_0$) to Fermi liquid formation ($x > x_0$), especially interesting are the lattice constant anomalies in $La_{2-x}Sr_xCuO_4$ (LSCO, $x_1 = 0.06$, $x_2 = 0.22$, $x_0 = 0.15$ - 0.16) that exhibit the internal elastic rigidity of the filaments, which stiffen the structure in much the same way that steel rods support reinforced concrete (a non-mean-field effect). This rigidity is exhibited in three ways: (1) through the positive bulge (~ 0.1%) in the planar lattice constants [22,11] above the values extrapolated from the regions $x < x_1$ and $x > x_2$; (2) from a positive bulge (~0.1%) in the temperature dependence below 150K of the mean square relative displacements determined by the correlated Debye-Waller factors $\sigma^2$ of the Cu-O bonds by EXAFS at $x = 0.1$, which is measurably weaker (< 0.03%) at $x = 0.2$ [23], and (3) ultraprecise measurements at $x = 0.15 \sim x_0$ ($T_c = 35$ K) of temperature-dependent positive (negative, in agreement with the Poisson ratio) bulges $\sim$ 0.0004 % of the planar (interlayer) lattice constants below $T_c$ [24].

Mean-field explanations of these effects encounter quantitative difficulties. For example, the signs of the shifts below $T_c$ are correctly predicted by a rigid band model [24], but the magnitude is much too small: about a factor of 10 too small, if one notes that the energy bands at the Fermi level are mixtures of Cu d and O p states. Perhaps these quantitative discrepancies could be resolved by a filamentary model, which would give larger magnitudes because of lowered dimensionality. Such a model would calculate changes in screening between the filamentary and Fermi liquid states; this lies outside the scope of this paper.

The most striking feature of the LSCO strain is that in the underdoped (purely filamentary) $x = 0.1$ sample the correlated $\sigma^2$ of the Cu-O bonds begins a superlinear increase (probably thermally activated, that is, exponential in $T^{-1}$) around $T_s \sim 110$K, with a midpoint near $T_m \sim 80$K [23]. This precursive effect is sometimes associated with superconductive droplets. Within the present model it is explained in terms of thermally activated dopant migration (especially near filamentary ends or weak links) that adjusts the states with an energy spacing $\sim W/N^{1/3}$, so that the Fermi level is pinned halfway between the highest occupied state and the lowest unoccupied state of each filament.

**7. Conclusions**

The self-organized (adaptive) dopant percolative filamentary model is entirely orbital in character; spin degrees of freedom are essentially irrelevant. This model has previously explained the origin of the intermediate phase *specifically* in the cuprates [1], and here it has explained the microscopic evolution of the electronic structure of that phase, as measured both by ARPES [2] and by time-resolved relaxation spectroscopy [9]. The latter technique was used in a number of previous experiments (see [9] for references), with results that were difficult to interpret topologically. By eliminating artifacts associated with heating, etc., and surveying the phase diagram carefully and completely, [[9] obtained results remarkably parallel to [2]. These results appear to be correct, and deserving of an equally correct theoretical model, which this paper has attempted to initiate.

It is important to realize that filamentary structure characterizes the intermediate phase even in apparently simple nonmagnetic cases, such as transmutation (randomly) doped semiconductor impurity bands, where there is also an intermediate phase whose existence was overlooked for decades in mean field theories [1]. Mean field theories do not "capture the essential physics" of metal-insulator transitions, which are essentially percolative connectivity transitions; instead, they mask the physics, often involve irrelevant spin degrees of freedom, and usually provide a misleading description of the microscopic interactions at all levels. The estimate given here of the ratio of relaxation times between the metallic and filamentary phases reflects averages, but only over nanodomain dimensions, not over optical wave lengths or, even worse, the entire sample.

*Postscript.* Plane wave models cannot consistently explain the observed d-wave anisotropy of the energy gap and the large magnitudes of $T_c$ [25], but the filamentary model does [14].

**Figure Captions**

Fig. 1. A sketch of the dopant-assisted percolative filamentary model for cuprates layered into predominantly metallic $CuO_2$ planes (also BiO planes in BSCCO) and semiconductive planes (here labeled SrO). This sketch is essentially unchanged since 1990 [4]. The metallic regions of the $CuO_2$ planes are indicated by wavy curves; these regions are separated by semiconductive nanodomain walls, indicated by straight lines. In (a) we have two such walls, marked 1,2. In wall 1 there is a dopant in the SrO layered labeled $D_1$. A coherent filamentary current path can go from A to B in the $CuO_2$ plane by utilizing $D_1$ as a bridge to bypass wall 1. The filament ends at B. In (b) a second dopant

$D_2$ has been added to wall 2, so that now the filament continues from A through B to C. In (c) $D_2$ has been replace by a surrogate excitonic bridge $E_2$; this structure explains the positive value of $\Delta R/R$ observed in underdoped samples in picosecond relaxation experiments [9]. Finally in (d) we have the overdoped case, where the exciton overlaps a dopant, locally destroying the coherence of the filament, and giving rise to a negative value of $\Delta R/R$.

Fig. 2. (a) A sketch of $Z(x,\alpha,E_F)$ (in arbitrary units, different for each curve) for $La_{2-x}Sr_xCuO_4$ (LSCO) [2] for x = 0.03, 0.15 (almost optimal doping), 0.18 and 0.22 (both overdoped); [2] also gives data for x = 0.05, 0.07, and 0.10. The central feature of these data, explained here, is the abrupt jump between x = 0.15 and x = 0.18 that is especially large near $\alpha = 0, \pi/2$, but is small near $\alpha = \pi/4$. This difference is illustrated in Fig. 2(b).

Fig. 3. For the reader's convenience the relaxation data from [9] are reproduced here: (a) $\Delta R(x)/R$ changes smoothly with x, and reverses sign at $x = x_0$; (b) $\tau(x)$ changes abruptly (essentially a step function) at $x = x_0$. The ps time scale of $\tau(x)$ implies that phonons play an essential part in the relaxation. Because the relaxation is essentially dielectric relaxation, spins are irrelevant.

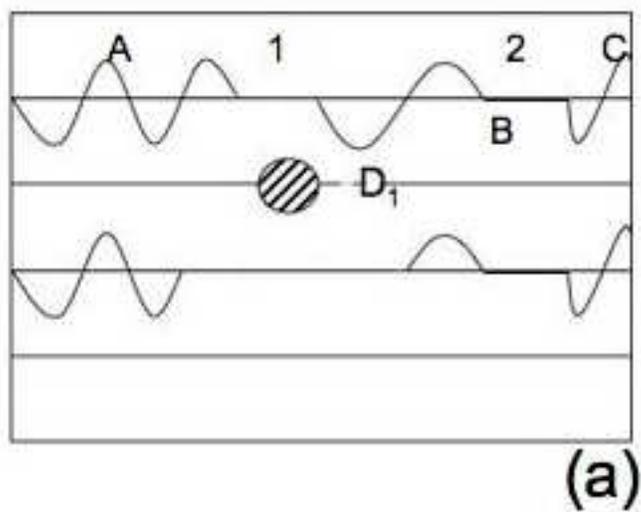
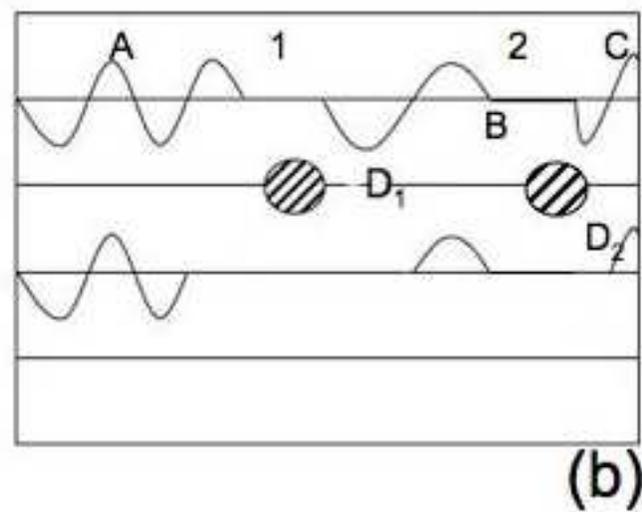
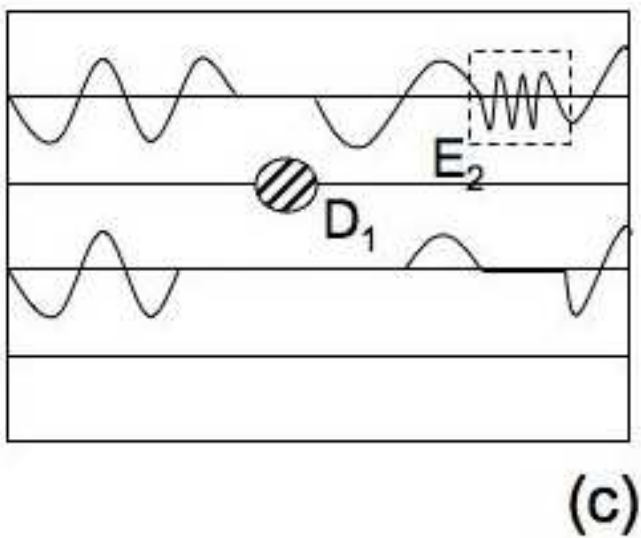
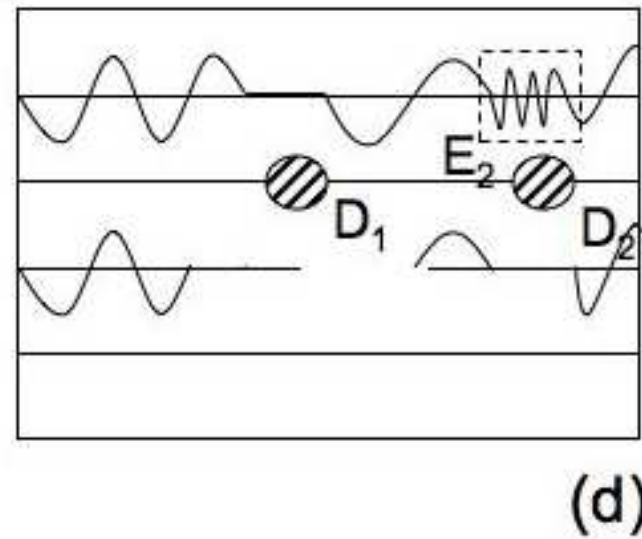

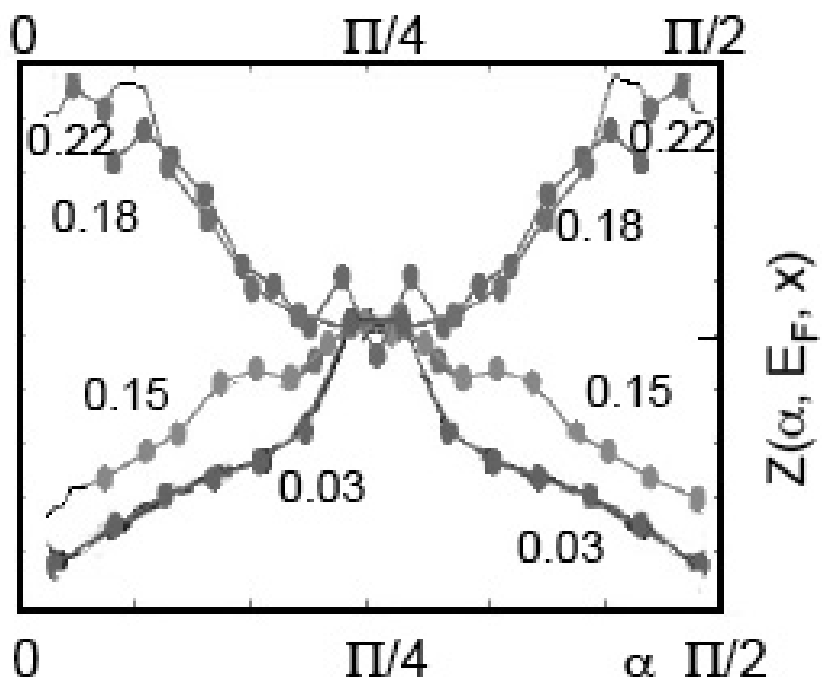

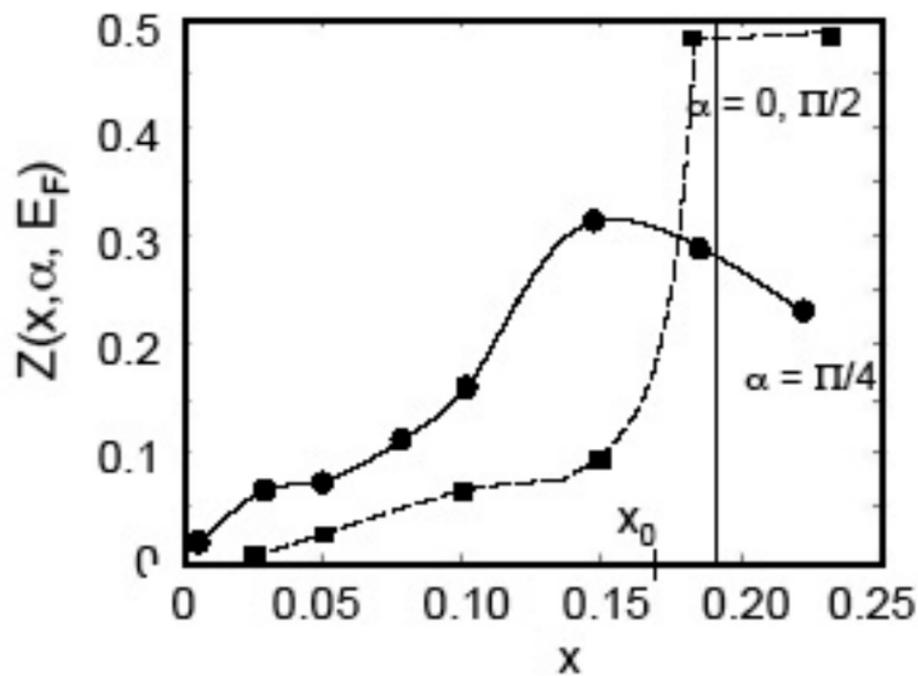

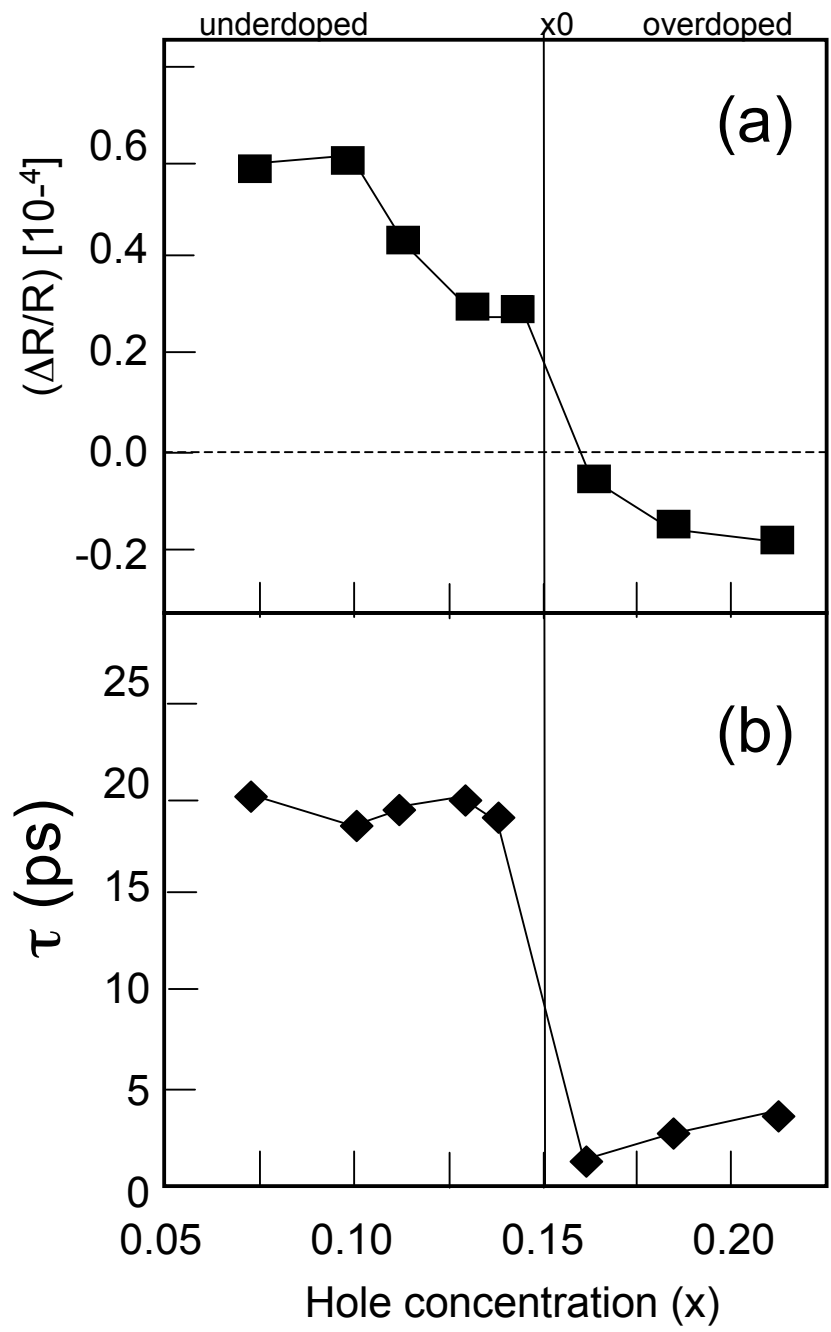